# Need for Mine Safety: An ICT based solution to make mines safer

*Dhruv Srivastava[1], Vaibhav Mehta[2], Prachi Chauthaiwale[3] and Dr. Priya Ranjan[4]*

1. *B.Tech. III Yr, Department of Electronics & Instrumentation, Indian School of Mines University, Dhanbad – 826004, India, E-mail: dhruv15125@ismu.ac.in*

2. *B.tech III Yr, Communication and Computer Engineering, LNM Institute of Information Technology, Jaipur, India, E-mail: vaibhav.08@lnmiit.com*

3. *B.Tech IV Yr, Department of Electronics & Communication, Visvesvaraya National Institute Of Technology, Nagpur, India, E-mail: prachi@ece.vnit.ac.in*

4. *Assistant Professor, Department of Electrical Engineering, Indian Institute of Technology, Kanpur – 208016, India, E-mail: ranjanp@iitk.ac.in*

# Contents:

# 1. <u>Abstract</u>

**Concern for the human security inside mines is as old as the mining itself. However, ICT (Information and communication technologies), which has impacted human life in so many ways has not been much used for making mines safer. We propose a method that has been practically implemented which can enhance mine safety enormously. It is based on integration of wireless sensor network with an external network through a gateway.**

# 2. <u>Background</u>

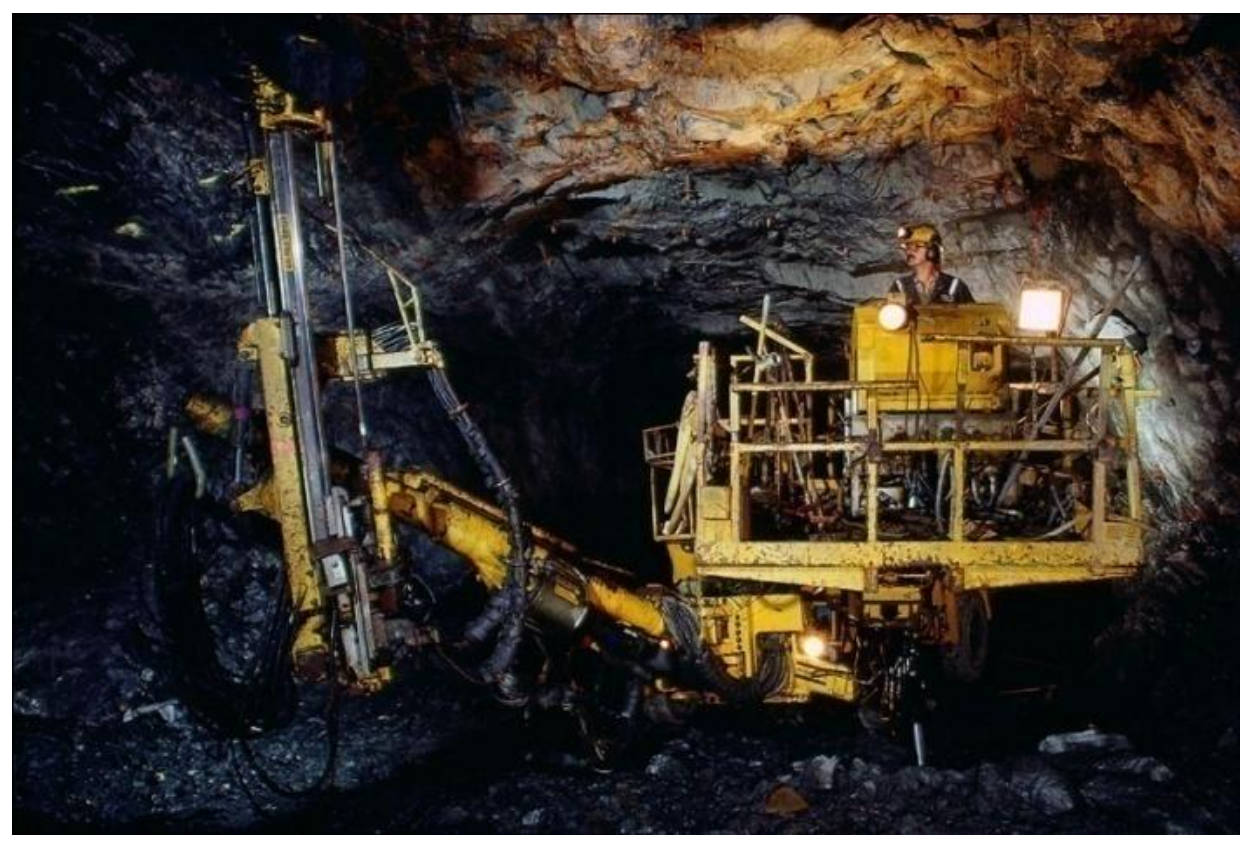

Imagine the world without coal, copper, uranium, iron-ore, gold and aluminum. We would have been in the era of early man even by now without these resources. All these metals as we use them or know them are extracted from the ores. The ores are naturally occurring minerals often buried deep inside the earth. These minerals are available to us because of the hard work put in by thousands of people working in the Mines. But have you ever thought of the kind of situation they work in? The reality is mind boggling.

Hundreds of people die every year in the mines due to lack of safety measures. Workers who work in the mines are always at a risk of death. The environment inside the mine is very difficult and apart from risk of losing lives, their exposure to harmful substances and the high probability of getting severe diseases is almost inevitable. **An average labor working in**

***Jharia( Dhanbad, India)* in the coal mines get paid at barely Rs. 50 per *bori* (15-20 kg) of coal that he mines out. Many children are also made to work in those terrible conditions for free.** And all this happens at the expense of their lives and risk of getting dreadful diseases. Astonishingly, things have been so terrible from the past few decades. Is it fair enough? Can human life be so cheap? Why cannot we change this deadly scenario?

## 3. Introduction: Need for Mining safety

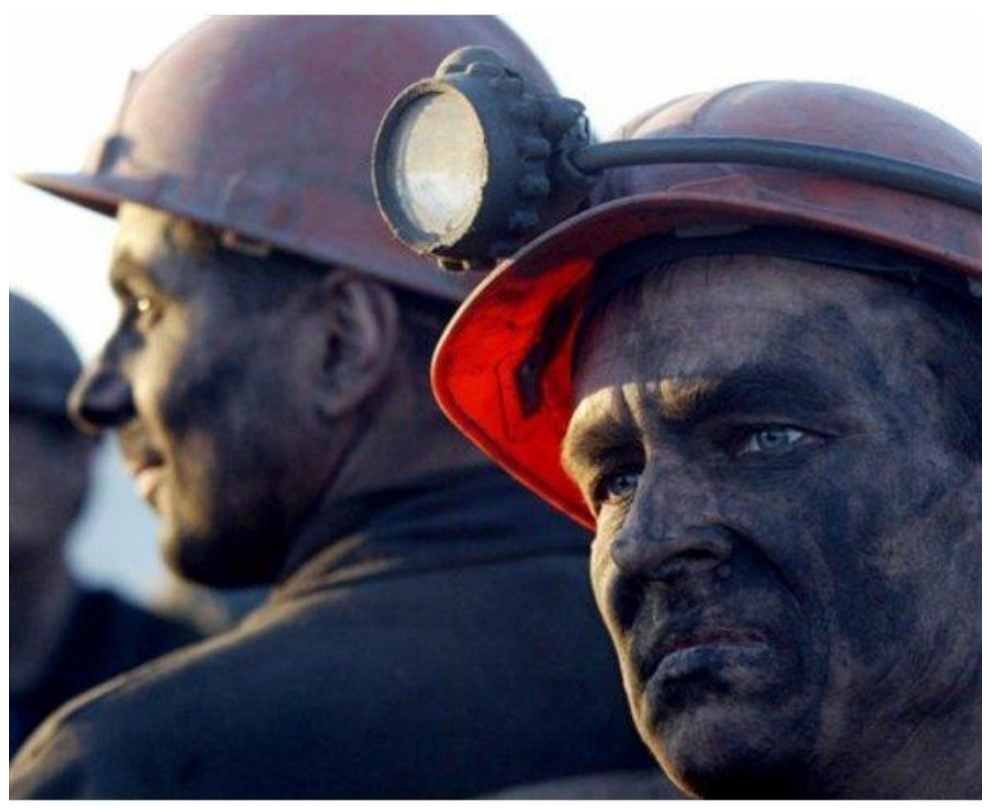

Thousands of people have lost their lives in Mines till date. History tells us about some huge mine disasters that have happened all around the world. Although, some of the disasters were not due to human mistakes but surely a consequence of mine safety measures. Why cannot we have better technology for mine safety issues? Advancements in technology are already taking at enormous rate, every day we come across some new gadget, services and applications being launched. Unfortunately, the uptake of technology is slow in the mining sector. So, why advancements in technology for Mine safety should be left out in a corner?

***Here are some shocking figures showing about the mine disasters that took place around the world (According to the documented statistics, the actual figures are much more and few are not being documented):***

### 3.1 SOME RECENT MAJOR MINING DISASTERS:

* April 2010 - USA - An explosion killed 29 miners in West Virginia in the deadliest U.S. mine disaster since 1984.

* February 2005 - CHINA - A gas explosion at the Sunjiawan colliery of state-owned Fuxin Coal Industry Group kills 214.

* November 2005 - CHINA - A gas explosion kills 169 people at state-owned Dongfeng coal mine in Heilongjiang province.

* September 2006 - INDIA - Fifty miners are killed after the roof of a coal mine collapses following an explosion in the eastern state of Jharkhand.

* September 2006 - KAZAKHSTAN - At least 41 people are killed after an underground explosion at Mittal's Lenin mine.

* March 2007 - RUSSIA - Blast rips through Siberian coal mine, killing at least 110 people.

* September 2007 - CHINA - Coal mine shaft floods in the eastern province of Shandong, killing 181 miners.

* September 2008 - CHINA - A mudslide caused by the collapse of a mine waste reservoir in northern China kills 254.

* November 2009 - CHINA - A gas explosion at a coal mine in northeast China kills 104.

* March 2010 - SIERRA LEONE - At least 200 people are killed when a trench collapses at a gold mine in Sierra Leone.

In ***Annexure 1*** we have given a statistics of major coal mine disaters happened over a period of time. The list is not exhaustive but surely provides glimpse of the enormity of the problem.

## 3.2 What makes the mines unsafe?

There are various things which make mines unsafe; e.g. mine ventilation, flooding of water, falling of roof, methane explosion, coal dust explosion etc. One of the major issues is the low level of oxygen, and concentration of noxious gases like carbon monoxide nitrogen, and methane. Their inappropriate concentration can cause serious damage to human beings present in the mines. There are chances of suffocation and even explosive conditions might arise. Most of the disasters which take place happen due to the explosion of methane gas and explosion of coal dust. If the decision makers who are located at considerable distance come to know the situation in real time they can initiate the corrective measures that can avoid accidents.

# 4. ICT based solutions to enhance mine safety

Several measures must be taken to improve and enhance the safety measures for mining. Focusing on ICT (Information and Communication technologies) based solutions; the web enablement of applications and other capabilities delivered over broadband communications systems can be used. Tele-robotics (operation of a machine at a distance), wireless technologies, radio-frequency identification (RFID) and global positioning system (GPS) technologies, which track the movement of minerals and equipment are few other features that may be added in mines.

## 4.1 Proposed Method

We propose a significant and efficient method for the real-time monitoring of various noxious gases in the mines and also parameters like temperature, moisture, water level etc. These are the parameters responsible for most of the mine accidents. The heart of this method is wireless sensor network's integration with an external network through a gateway. By this method, data about the above mentioned parameters can be accessed expediently from long-distance. We can also put up data about the parameters on a common server or can be updated on a website. And many other applications can be made.

### 4.2 How to implement the method?

We can prepare a wireless sensor network and integrate with an external network like intranet/ internet or any other telecom application through a gateway. The wireless sensor network must be planted in a mine. This network should have several nodes placed at various locations of the mine. Each node should be planted at different locations of the mine covering the whole area. Each node needs to be fabricated with temperature sensors like TMP-275 and various gas sensors like TGS-2611, TGS-2442, and TGS-2600 which can measure the parameters like temperature, concentration of oxygen, carbon monoxide nitrogen, and methane.

The whole network needs to be set in a suitable network topology preferably a hybrid of ring and star topology. Though, the topology to be selected depends on the basis of the area to be covered, data redundancy issues or energy optimization. But here we are interested in the coverage of more area. The network has cluster heads which collects data from every leaf node and then these cluster heads sends the whole data to the base station which is connected to an external network through a gateway.

## 4.3 Effective topologies for the wireless sensor network

***Figures 4.3.1, 4.3.2 & 4.3.3 are few of the most effective topologies which can be used for the wireless sensor network depending upon the requirement.***

***Terminology used:***

***i.*** *Interrupt call*- When a node sends a message to another node asking to send the data it has.
***ii.*** *Flow of data*-Sending information from node to another i.e. wireless communication between the nodes.
**iii.** *Topology*- The way/structure in which the network has been set up.

***Symbols used:***

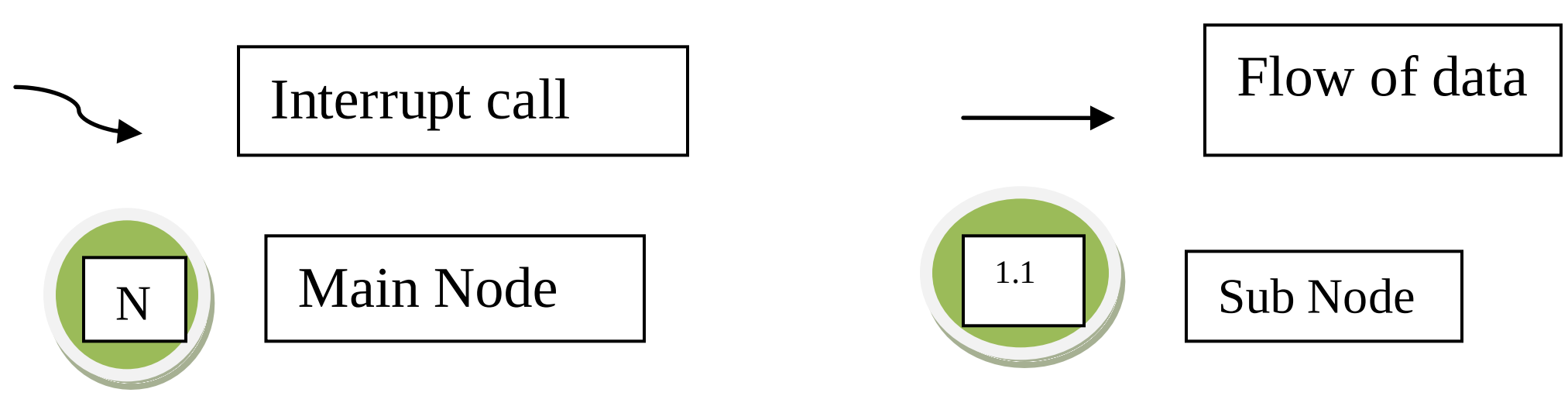

***Assumption:*** Wireless range of every node is same.

### 4.3.1 **Star on ring topology :** Preferable when more distance coverage is needed.

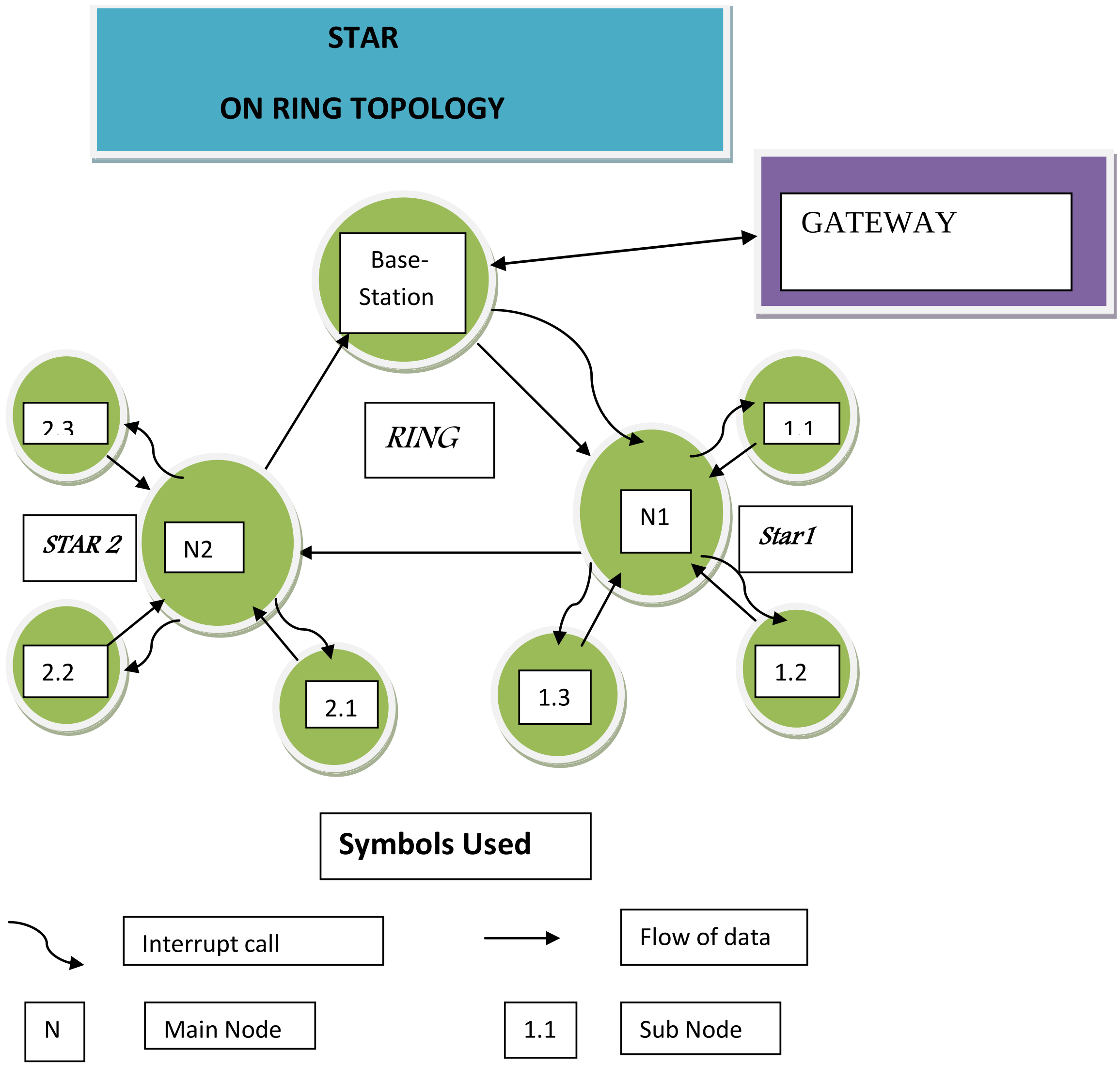

***Figure 4.3.1***

**Key Features of Star on ring topology**:

i. Whenever distance coverage needs to be increased, nodes (having sub- nodes) can be easily added to the main ring.
ii. Each main node carries information of its every sub-node.
iii. If any of the main nodes fail the whole network will fall.
iv. Data redundancy takes place to a certain level.

**4.3.2 Star on star topology:** To avoid data redundancy & faster communication

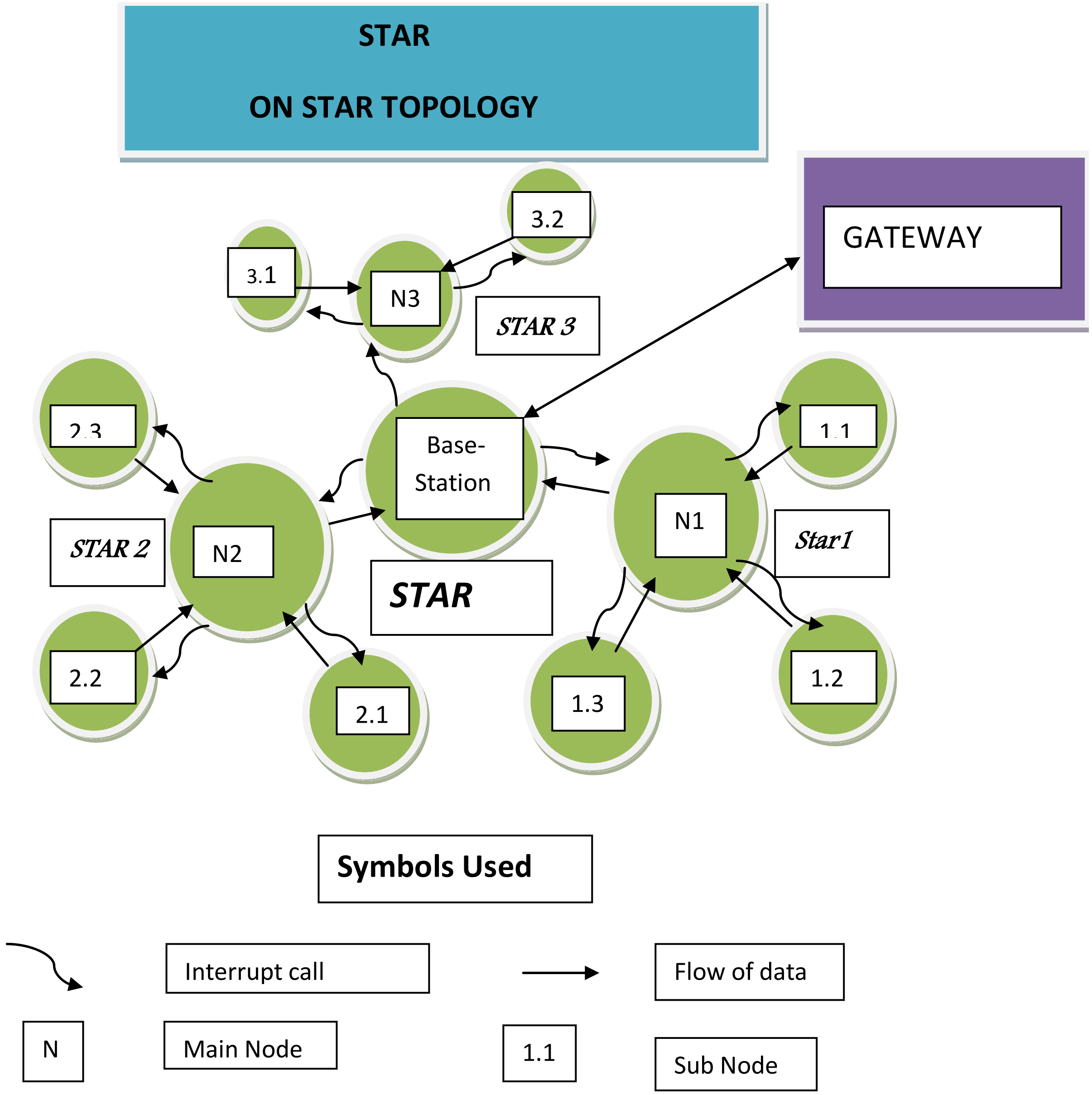

***Figure 4.3.2***

**Key Features of Star on ring topology**:

i. Data is collected and passed on very efficiently without data redundancy.

ii. Each main node carries information of its every sub-node.

iii. Even if any of the main nodes fail the whole network will keep functioning.

iv. Not suitable if more distance needs to be covered.

## 4.3.3 Tree Topology

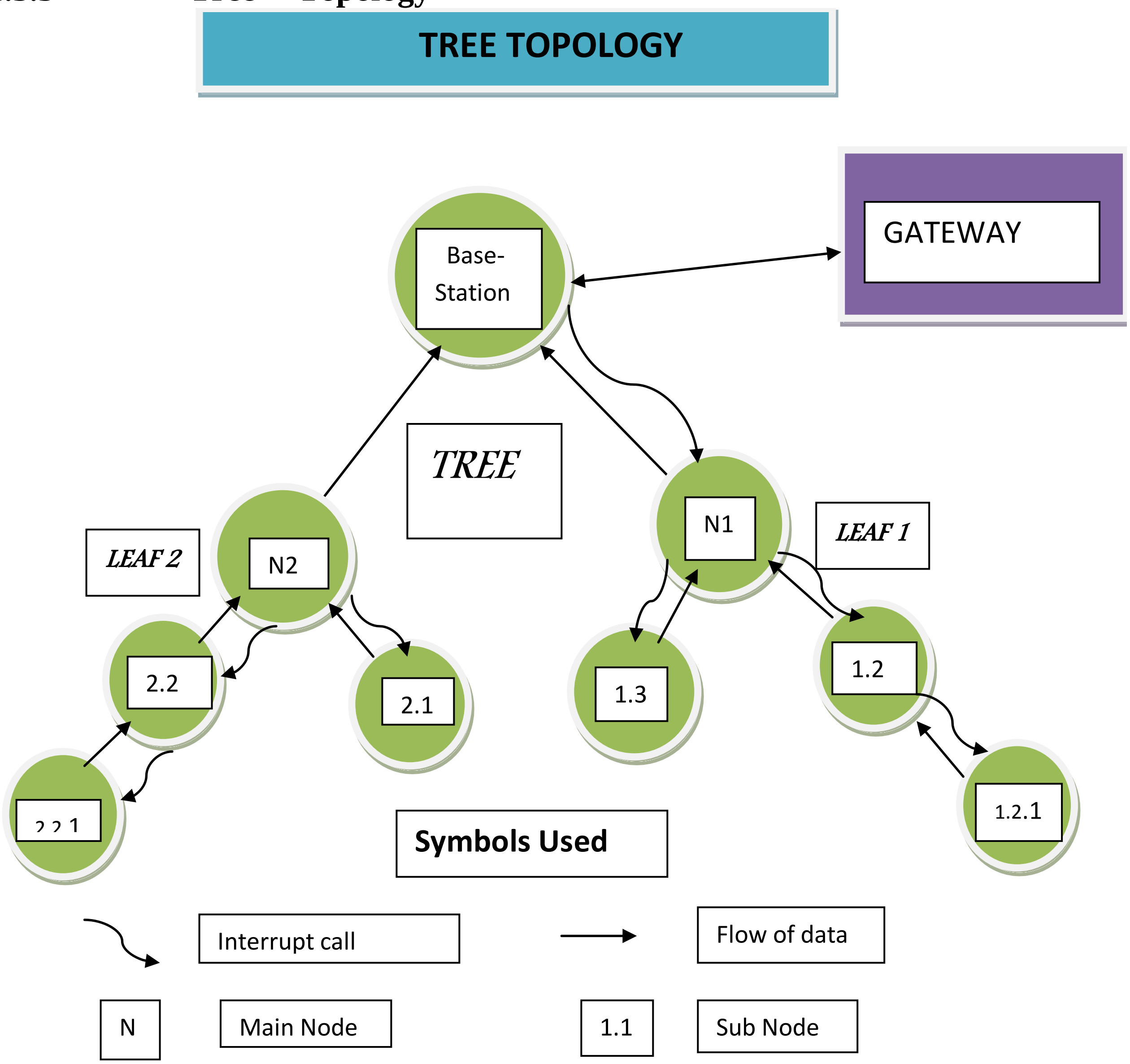

***Figure 4.3.3***

**Key Features of Tree topology**:

i. Data is collected and passed on very efficiently without much data redundancy.

ii. Each leaf carries information of its every leaflet.

iii. If a leaf fails then data of the leaflets cannot reach the base station.

## 4.4 Details of the Experiment:

### 4.4.1 Software used and required:

i. **AVR Studio 4:** For programming & debugging the AVR microcontroller.

**ii.** **X-CTU:** Allows the ability to program the radios' firmware settings via a graphical user interface.(Windows based)

iii. **MATLAB** (optional)

iv. **Docklight** (optional): To view the data coming from the serial port interface.

### 4.4.2 Hardware used:

**1.** **Indriya CS-03A14 Kit :** Wireless sensor module having options to plug in different combinations of sensors

2. **AVR Atmega 128L microcontroller**: 7.3728 MHz, 128KB flash, 4KB RAM processing unit

3. **Temperature Sensor (TMP-275)**: $0.5^0$C accuracy digital response

4. **Ambient Light Sensor (APDS-9300)**: approximate human eye 16bit I2C compatible response

5. **XBee Wireless Radio (2.4 Ghz):** IEEE802.15.4 compliant radio with 30m-100m range

6. **USB-UART Interface (FTDI chip)**: For communication of hardware with a computer through serial communication

### 4.4.3 Description of the hardware used:

**1. Indriya CS-03A14**

Indriya CS-03A14 is an ambient wireless sensor module having options to plug in different combinations of sensors like Light, temperature, gas, Humidity, Air Quality sensors, Occupancy Detection etc. These Modules are powered with Indrion's low power routing algorithms. INDRIYA can be run from AC supply as well as battery, this adds to the flexibility in placement of sensors. It has a microcontroller AVR's Atmega 128L which is the brain of the kit where the processing of the data takes place. It also has a USB-UART interface for easy communication between the hardware and a computer through serial communication.

Key components fabricated on Indriya CS-03A14 are AVR Atmega 128L microcontroller, Temperature Sensor (TMP-275) and Light Sensor (APDS-9300).

***Figure 4.4.1*** *is a picture of the hardware we have used.*

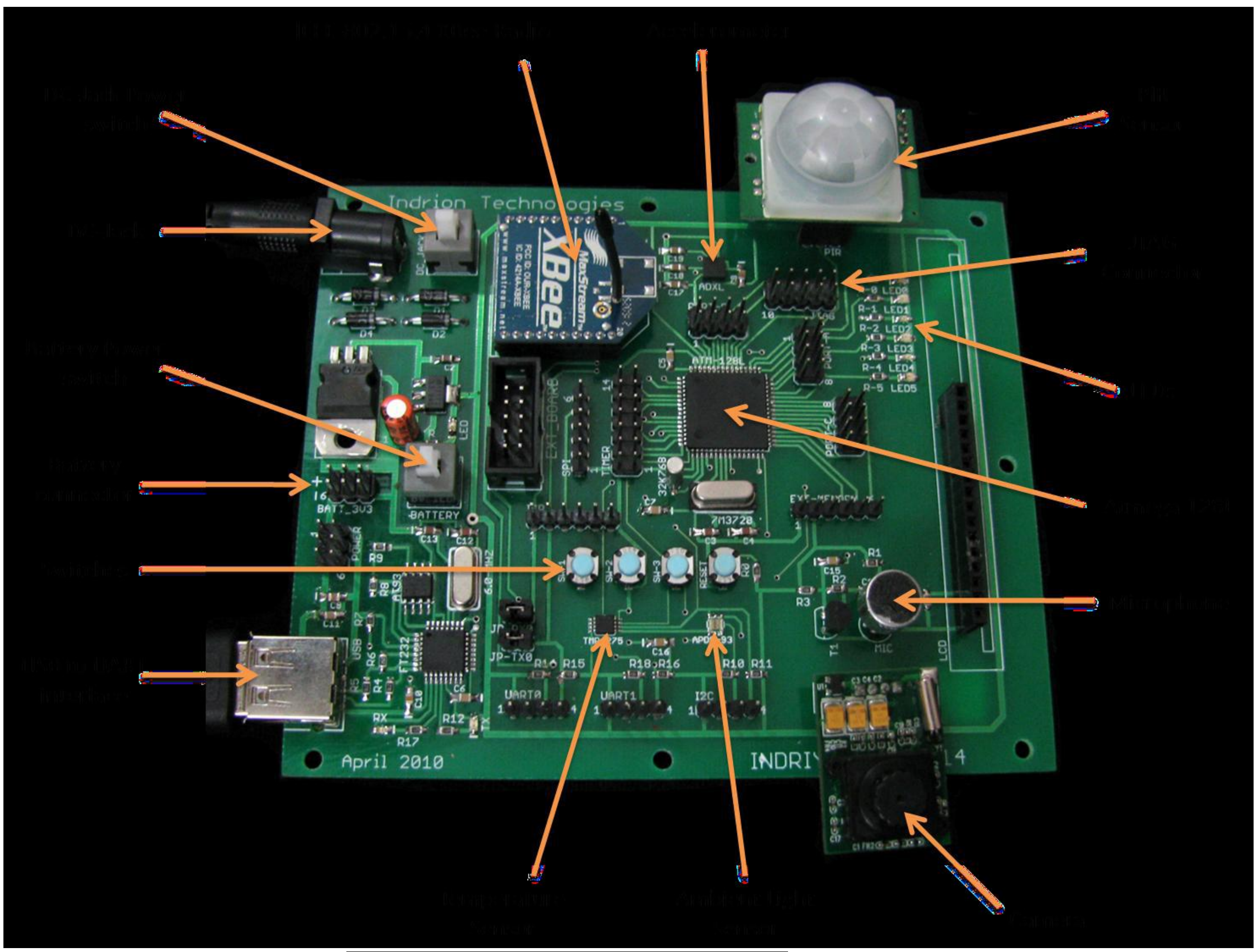

***Figure 4.4.1***

***Figures 4.4.2 & Figure 4.4.3*** *are schematic diagram & circuits of every component:*

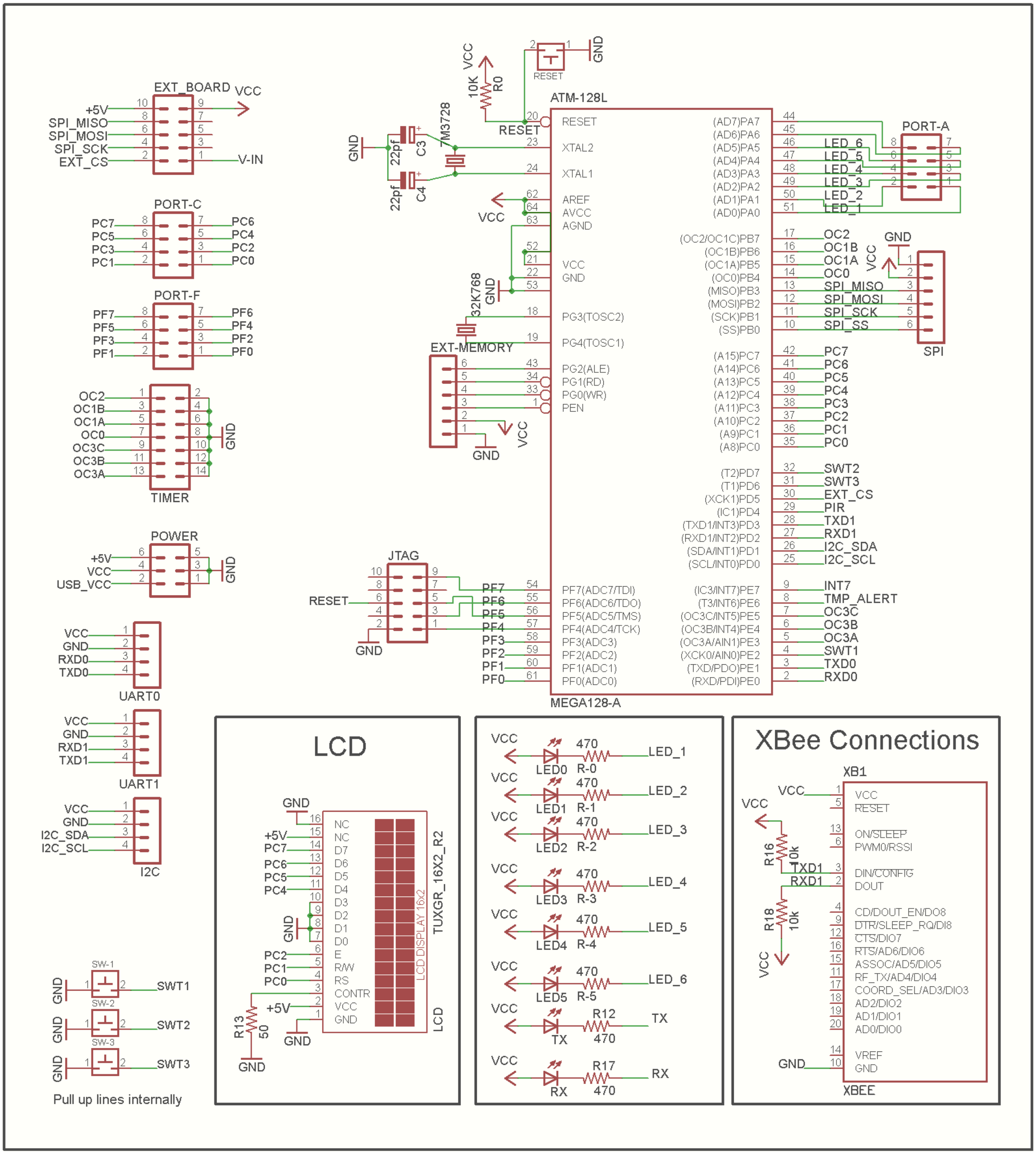

**Figure 4.4.2**

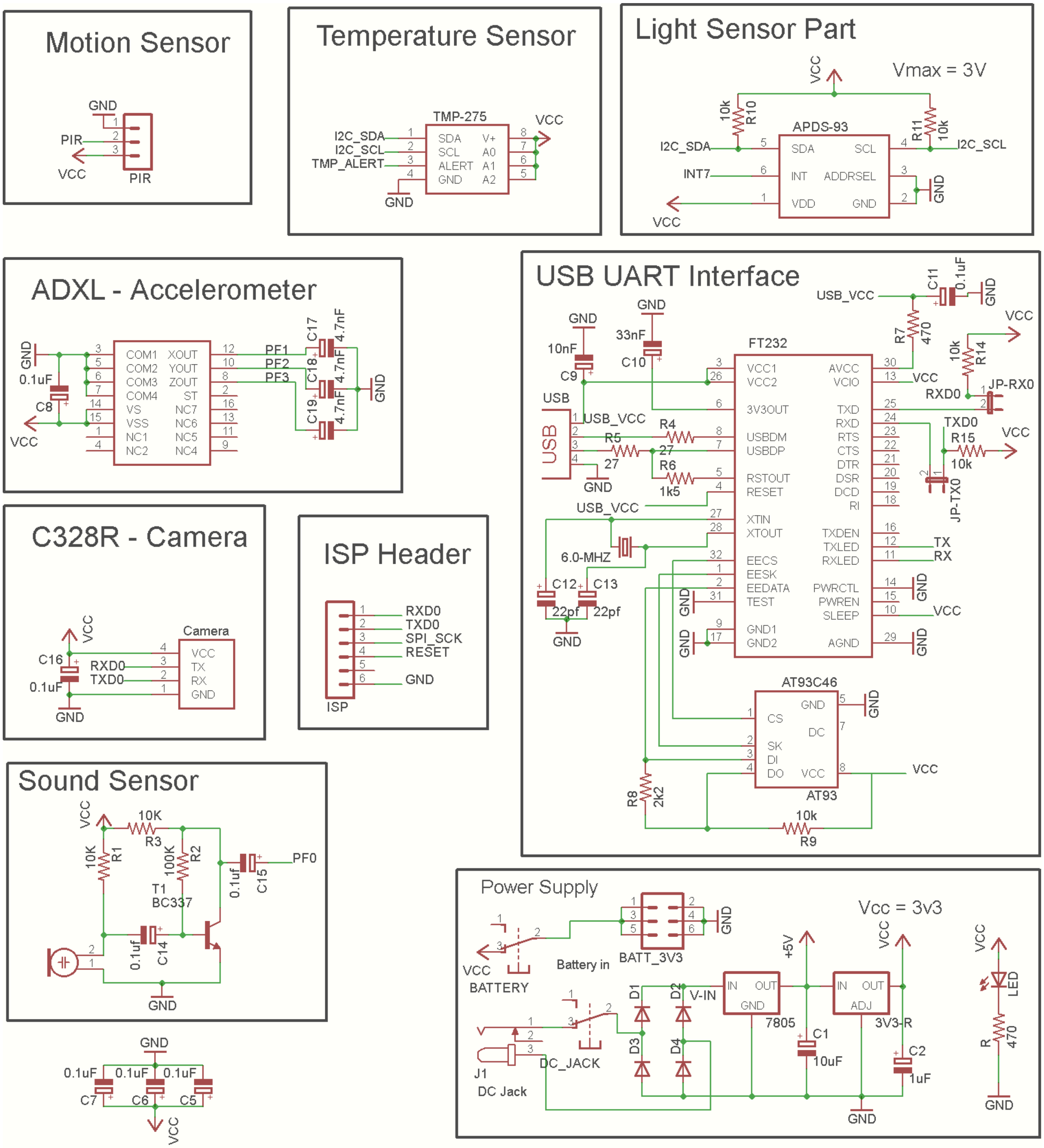
Motion Sensor
GND
PIR
VCC
PIR
Temperature Sensor
TMP-275
VCC
I2C_SDA
I2C_SCL
TMP_ALERT
SDA V+
SCL A0
ALERT A1
GND A2
GND
Light Sensor Part
VCC
Vmax = 3V
10k R10
R11 10k
APDS-93
I2C_SDA
I2C_SCL
INT7
SDA SCL
INT ADDRSEL
VDD GND
GND
VCC
ADXL - Accelerometer
GND
0.1uF
C8
VCC
COM1 XOUT
COM2 YOUT
COM3 ZOUT
COM4 ST
VS NC7
VSS NC6
NC1 NC5
NC2 NC4
PF1
PF2
PF3
C17 4.7nF
C18 4.7nF
C19 4.7nF
GND
USB UART Interface
USB_VCC
C11 0.1uF
GND
VCC
R7 470
GND
10nF
C9
GND
33nF
C10
FT232
USB
USB_VCC
R4
R5
27
27
R6
1k5
GND
USB_VCC
6.0-MHZ
C12 C13
22pf 22pf
GND
VCC1 AVCC
VCC2 VCIO
3V3OUT TXD
RXD
USBDM RTS
USBDP CTS
DTR
RSTOUT DSR
RESET DCD
RI
XTIN
XTOUT TXDEN
TXLED
EECS RXLED
EESK
EEDATA PWRCTL
TEST PWREN
SLEEP
GND1
GND2 AGND
10k R14
JP-RX0
RXD0
TXD0
R15
10k
VCC
JP-TX0
TX
RX
GND
VCC
GND
AT93C46
GND
CS
DC
SK
DI
DO VCC
AT93
VCC
R8 2k2
10k
R9
C328R - Camera
VCC
Camera
C16
0.1uF
RXD0
TXD0
GND
VCC
TX
RX
GND
ISP Header
RXD0
TXD0
SPI_SCK
RESET
GND
ISP
Sound Sensor
VCC
10K
R3
10K R1
100K R2
T1
BC337
0.1uf C15
PF0
0.1uf C14
GND
GND
0.1uF 0.1uF 0.1uF
C7 C6 C5
VCC
Power Supply
Vcc = 3v3
VCC
BATTERY
Battery in
BATT_3V3
GND
+5V
VCC
VCC
D1 D2
V-IN
IN OUT
GND
7805
C1
10uF
IN OUT
ADJ
3V3-R
C2
1uF
LED
R 470
GND
GND
DC_JACK
D3 D4
J1
DC Jack

**Figure 4.4.3**

## 2. AVR Atmega 128L microcontroller

The ATmega128 is a low-power CMOS 8-bit microcontroller based on the AVR enhanced RISC architecture. It has 128K Bytes of In-System Self-programmable Flash program memory, 4K Bytes EEPROM, 4K Bytes Internal SRAM, 53 Programmable I/O Lines and allows programming of Flash, Timers, I2C, SPI, ISP, Fuses and Lock Bits through the JTAG Interface**.**

➢ Main Peripheral Features:

- Dual programmable serial 2 USART's (Universal Synchronous Asynchronous Receiver/Transmitter)
- 8 Channel,10 bit ADC

Further technical details of Atmega 128L can be found in its datasheet (***Refer*** www.atmel.com/atmel/acrobat/doc2467.pdf***)***.

## 3. Temperature Sensor (TMP-275)

TMP275 is a temperature sensor from TI. It has a resolution of 0.5 degree centigrade. It is a digital output sensor interfaced with microcontroller via I2C interface.
Further technical details of TMP275 can be found at http://focus.ti.com.cn/cn/general/docs/lit/getliterature.tsp?genericPartNumber=tmp275&fi.

## 4. Light Sensor (APDS-9300)

Light sensor APDS-9300 is from Avago Technologies. The APDS-9300 is a low-voltage Digital Ambient Light Photo Sensor that converts light intensity to digital signal output capable of direct I2C interface. Each device consists of one broadband photodiode (visible plus infrared) and one infrared photodiode.
Further technical details of APDS-9300 can be found at: www.avagotech.com/docs/AV02-1077EN.

## 5. XBee Wireless Radio (2.4 Ghz):

This is the most important of the components for establishing wireless communication among the nodes.XBEE module is an IEEE802.15.4 compliant radio from Maxstream,Inc. It has a data rate of 250Kbps and range of 30m (indoor) and 100m (outdoor). It is interfaced with the microcontroller via UART.
Key features supported are:

- Source/Destination Addressing
- Unicast & Broadcast Mode
- Point to point, point to multi point & peer to peer topology supported
- Two Command Modes: Transparent & API (Application programming interface)

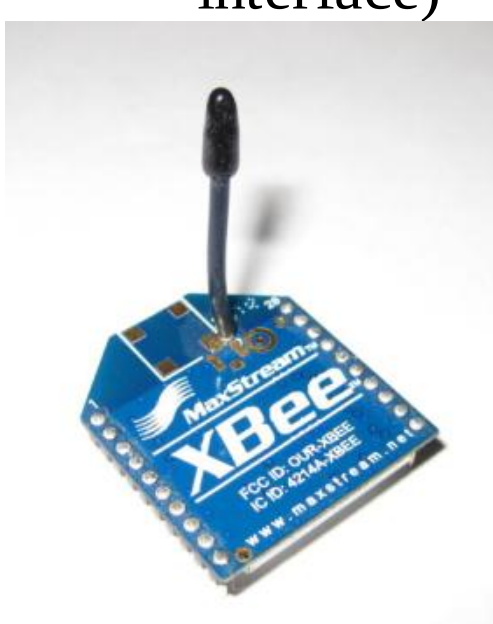

***Figure 4.4.4***

Further technical details can be found in its datasheet (***Refer*** [http://www.digi.com/products/wireless/point-multipoint/xbee-series1-module.jsp#docs](http://www.digi.com/products/wireless/point-multipoint/xbee-series1-module.jsp#docs)*)*.

## 6. USB-UART interface

It provides a USB interface for serial communication test applications. This USB interface is actually an UART- USB interface based on FTDI232 chip for PC to Indriya communication. All USB protocols are handled in the chip, so no knowledge of USB is required. The 232BM is entirely state machine based and no firmware is required. Support baud rates of up to 3 MB/second. The interface is as shown in figure below:

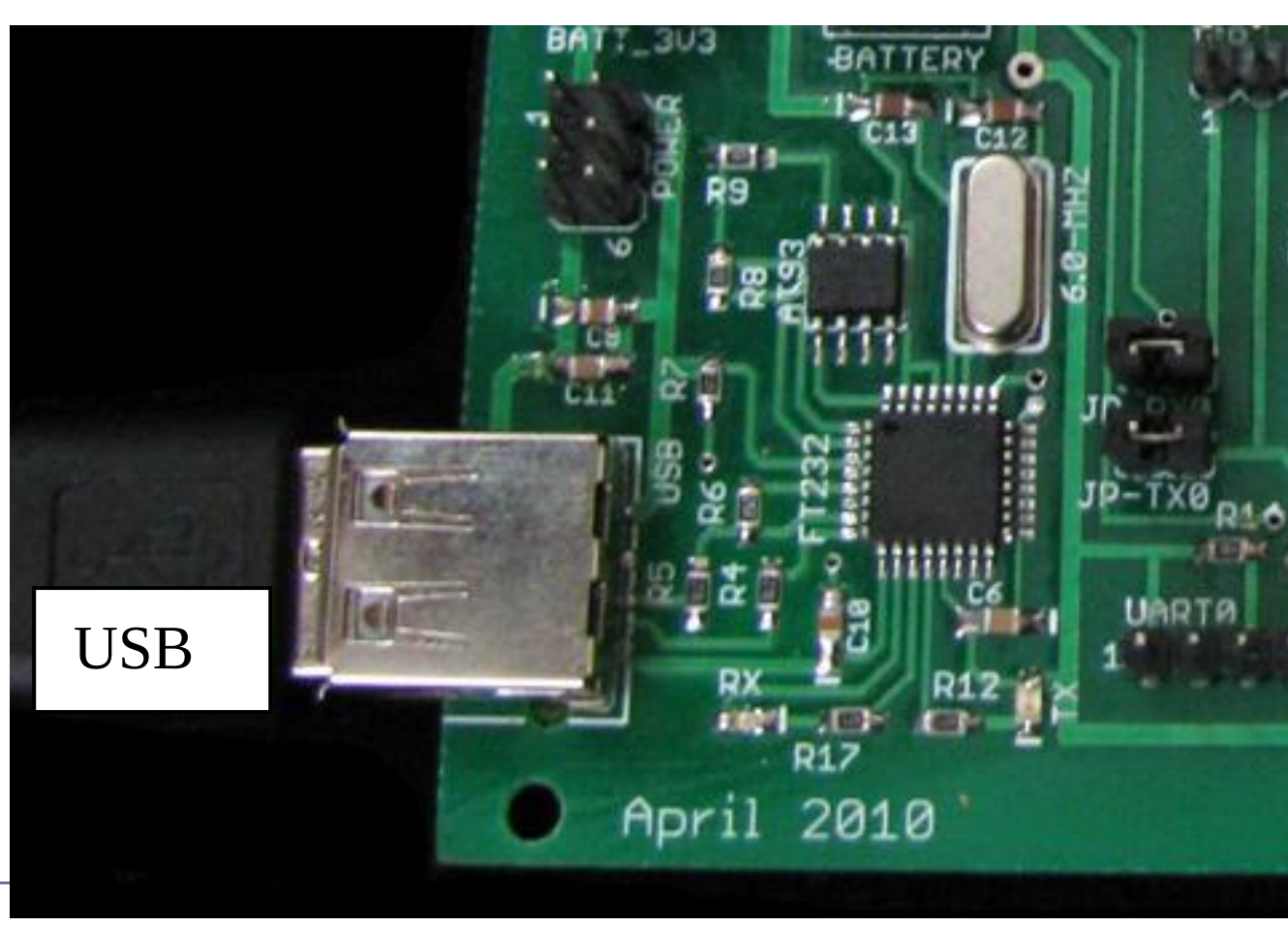

FTDI232 chip

***Figure 4.4.5***

## 4.5 Results and Discussions

### 4.5.1 Temperature display on www.ijaal.org

We made use of INDRIYA CS-03A14 kit components embedded on it like AVR's Atmega 128 microcontroller (brain), temperature sensor TMP-275. We successfully displayed the live temperature (with time stamping) of the laboratory on the website www.ijaal.org.

We programmed Atmega 128 which is a low power CMOS 8 bit microcontroller having AVR RISC architecture with the help of AVR studio software in C to work as a temperature sensor. The temperature sensor sends analog data to the microcontroller through I2C communication which then forwards it to the Universal Asynchronous Receiver Transmitter (UART) through FTDI communication. Then entire data is being sent to the computer serially (with the help of USB-UART interface) i.e. through serial port interface. The serial port interfacing has been done in C for linux based systems to make the work **open source**. After reading the data through serial port the data is processed and is converted to the closest value of the temperature. The program made keeps on running in an infinite loop and continuously updates the temperature of the lab with time stamping in a file. Then finally this file is being uploaded on the ijaal server using Ajax and you can see the temperature display on www.ijaal.org.

We can envision a fully fledged technical program leveraging state of art wireless sensor network technologies towards an intelligent and secure mining environment. It can render the workplace of miners safe from unpredictable situations and accidents which lead to loss of human lives and prosperity.

### 4.5.2 Implementation of multi-hop network of sensor nodes & integration with an external network

We have implemented a multi-hop network of sensor nodes which are capable of monitoring temperature, light intensity and other ambient data. The base station for a particular deployed network is responsible for aggregation of data from all nodes and sending it to the main server via a gateway. The server can provide service to multiple clients. A customized application layer has been built on the client machines to facilitate efficient data monitoring and cluster selection from a deployed WSN.

We have used a tree topology for the wireless sensor network having 2 main leaves (cluster heads) & their 2 leaflets (sub-nodes) each. In total we monitored the values of temperature & light intensity of 6 nodes in total. ***Figure 4.5.1 shows a flow chart of what exactly have we implemented.*** We also plotted graphs for both the readings i.e. temperature & light intensity coming from every node. The graphs being plotted at the time of the experiment can be seen ***Section 4.6.***

The readings of temperature & light intensity coming from every node were written onto a file. The file was being updated at regular interval of time and was sent to different clients on their request, respectively.

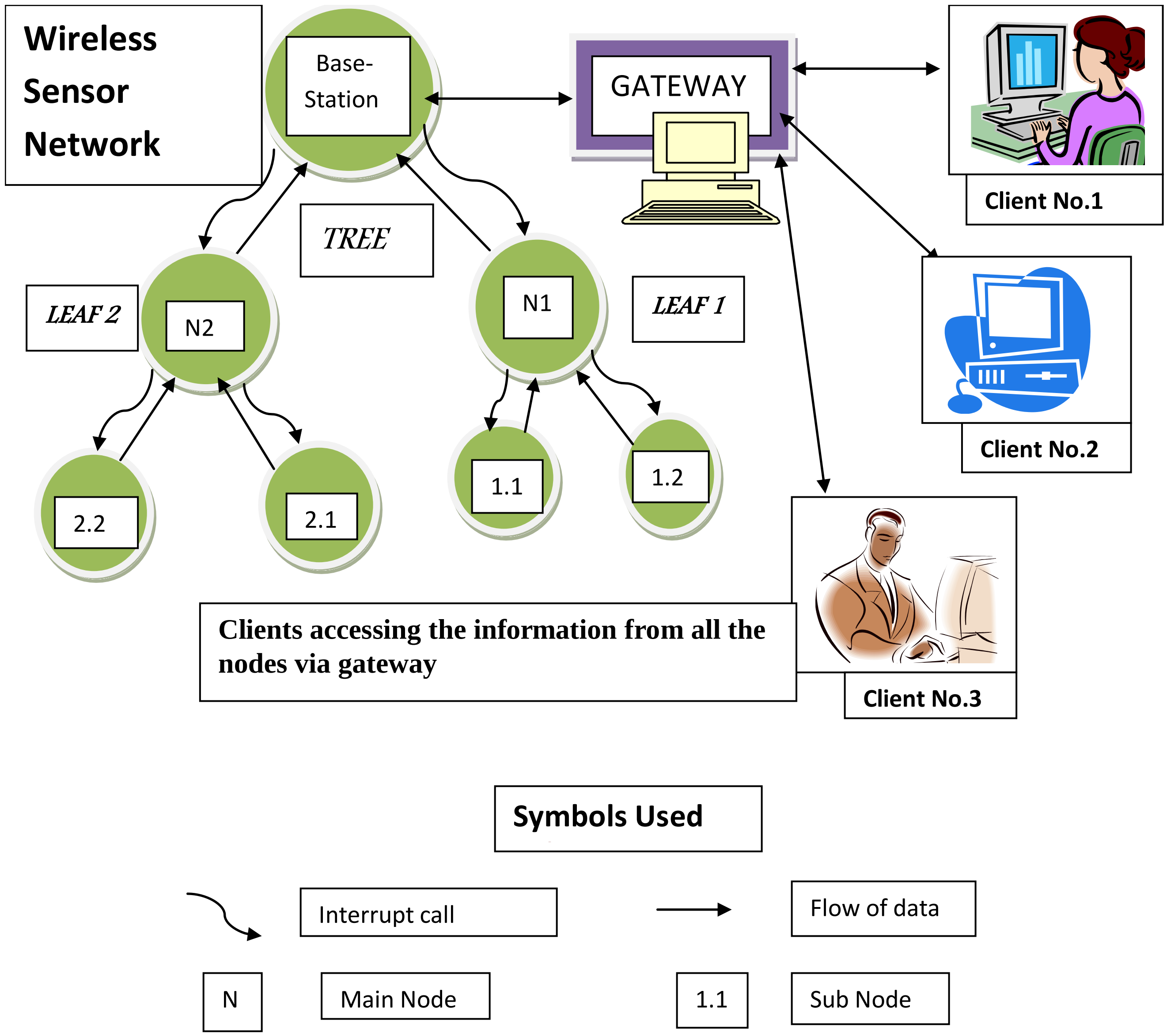

*Figure 4.5.1*

***Figure 4.5.1**: A flow chart of what exactly have we implemented:*

## 4.6 Demonstration of our achieved results

### 4.6.1 Temperature display on www.ijaal.org

We made use of INDRIYA CS-03A14 kit components embedded on it like AVR's Atmega 128 microcontroller (brain), temperature sensor TMP-275. We successfully displayed the live temperature with time as detected by the sensor of the laboratory on the website www.ijaal.org.

### 4.6.2 Graphical display of temperature & light intensity of every node in the sensor network

The readings of temperature & light intensity coming from every node placed at various location of the laboratory were taken. The readings taken were then plotted on graphs which are shown in ***figures 4.6.1-4.6.6***.

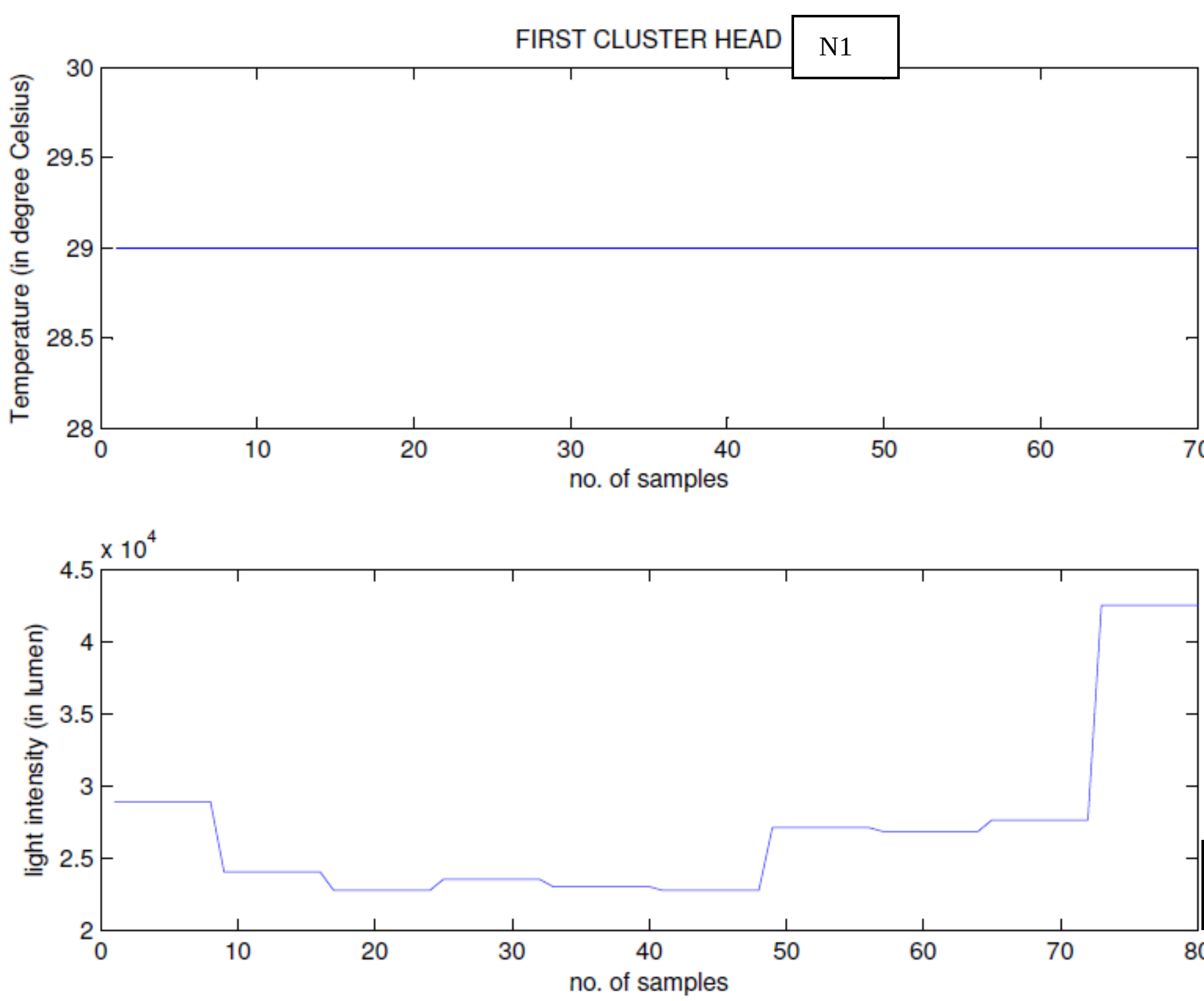

***Figure 4.6.1***

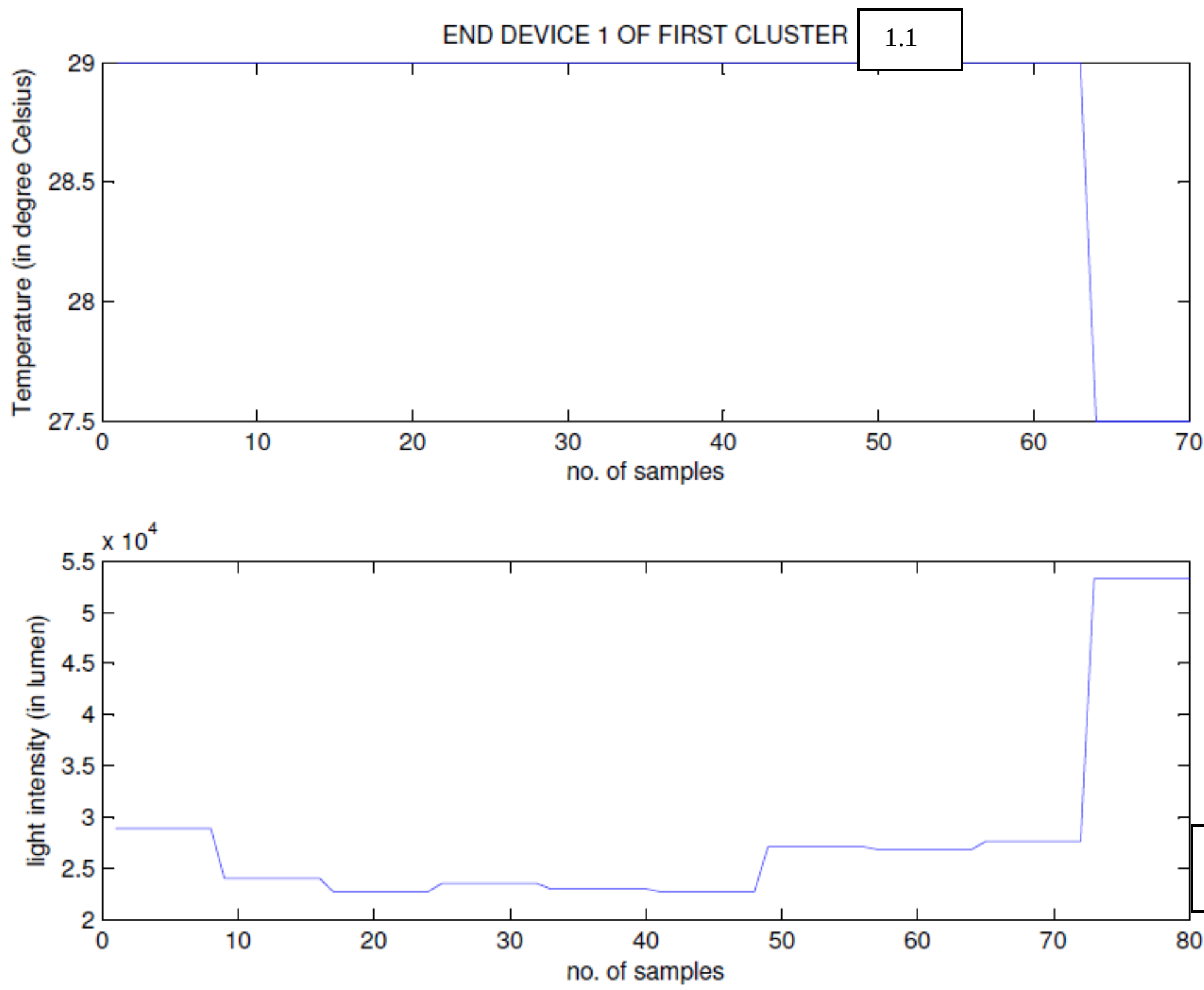

***Figure 4.6.2***

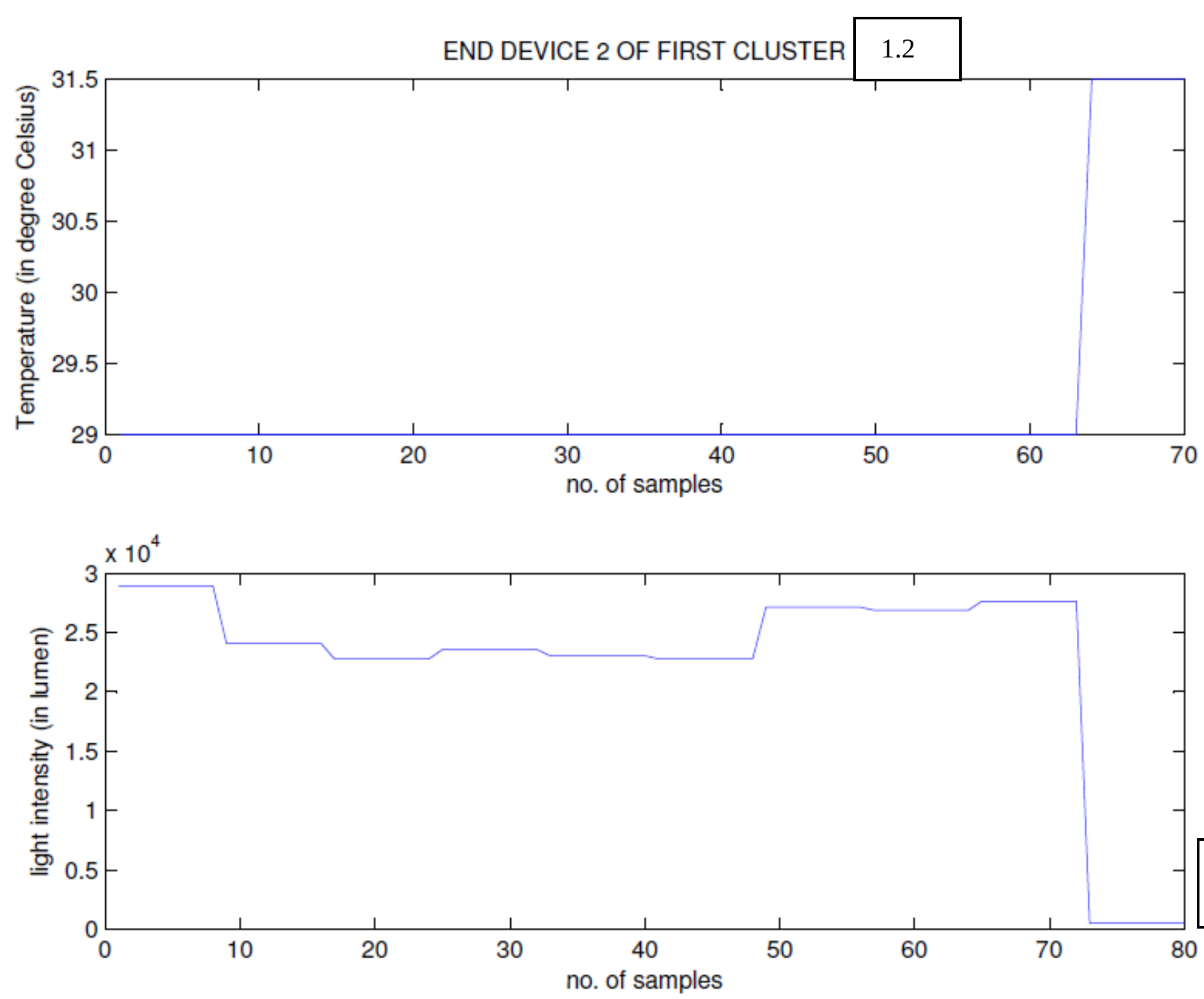

***Figure 4.6.3***

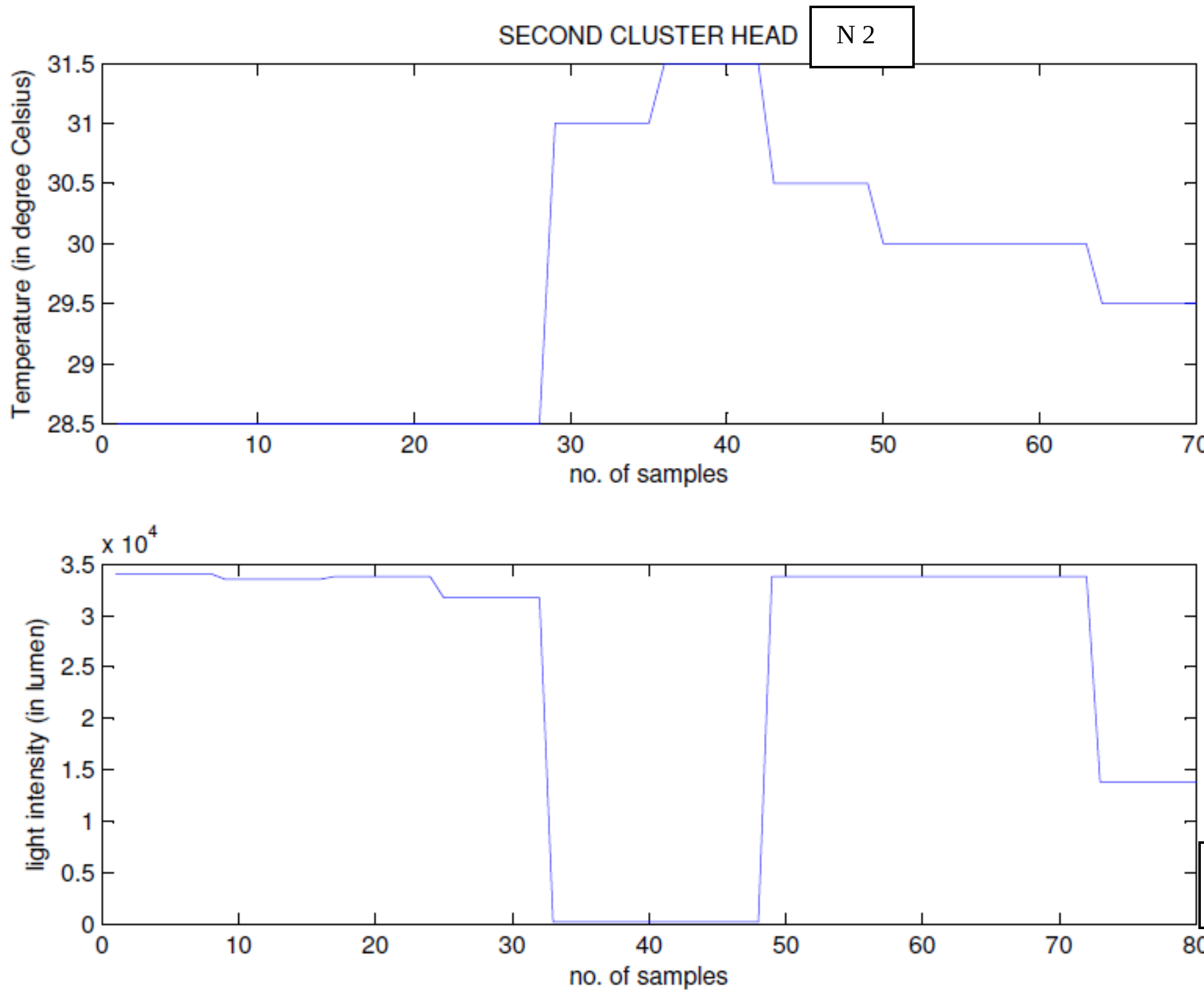

***Figure 4.6.4***

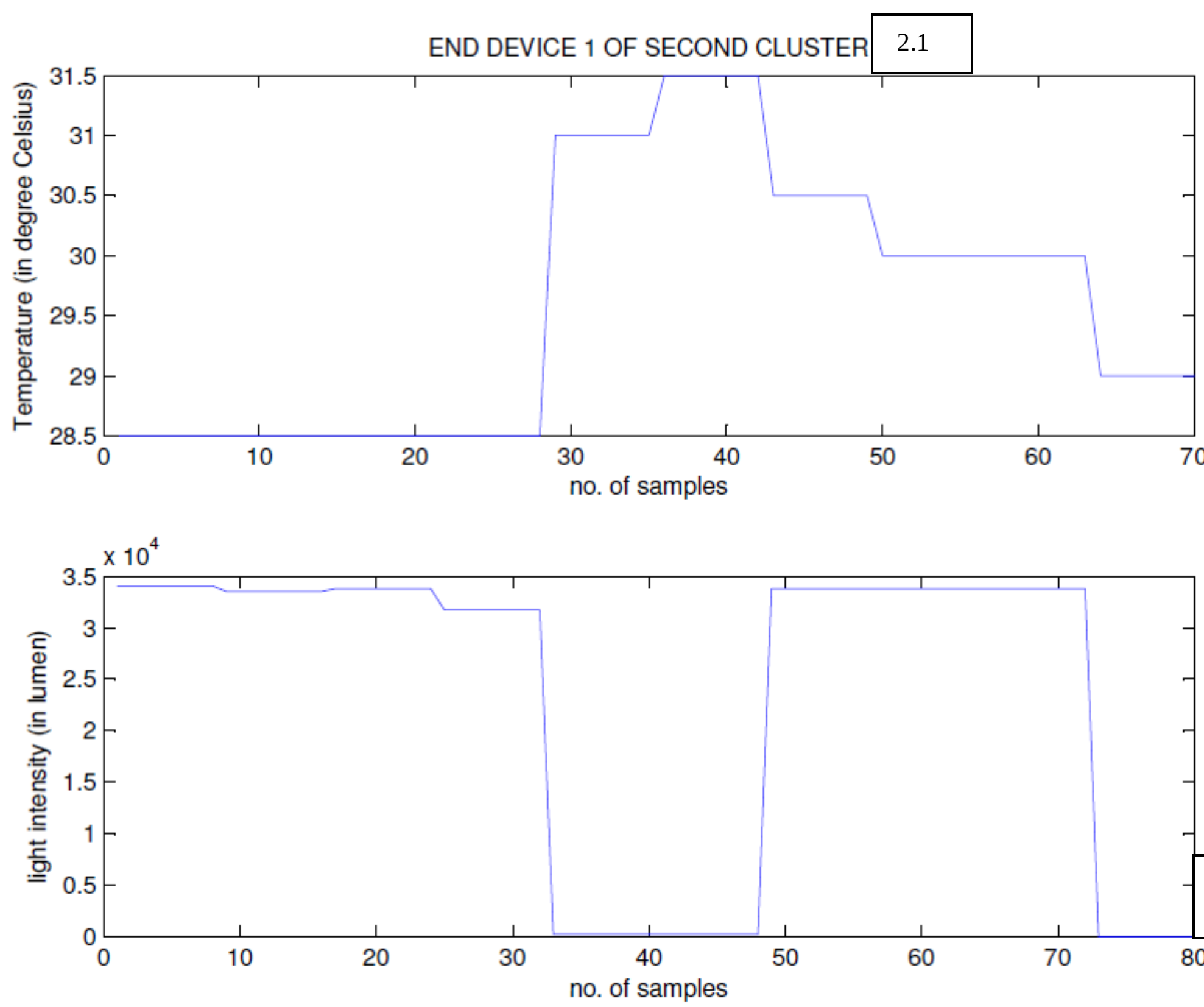

***Figure 4.6.5***

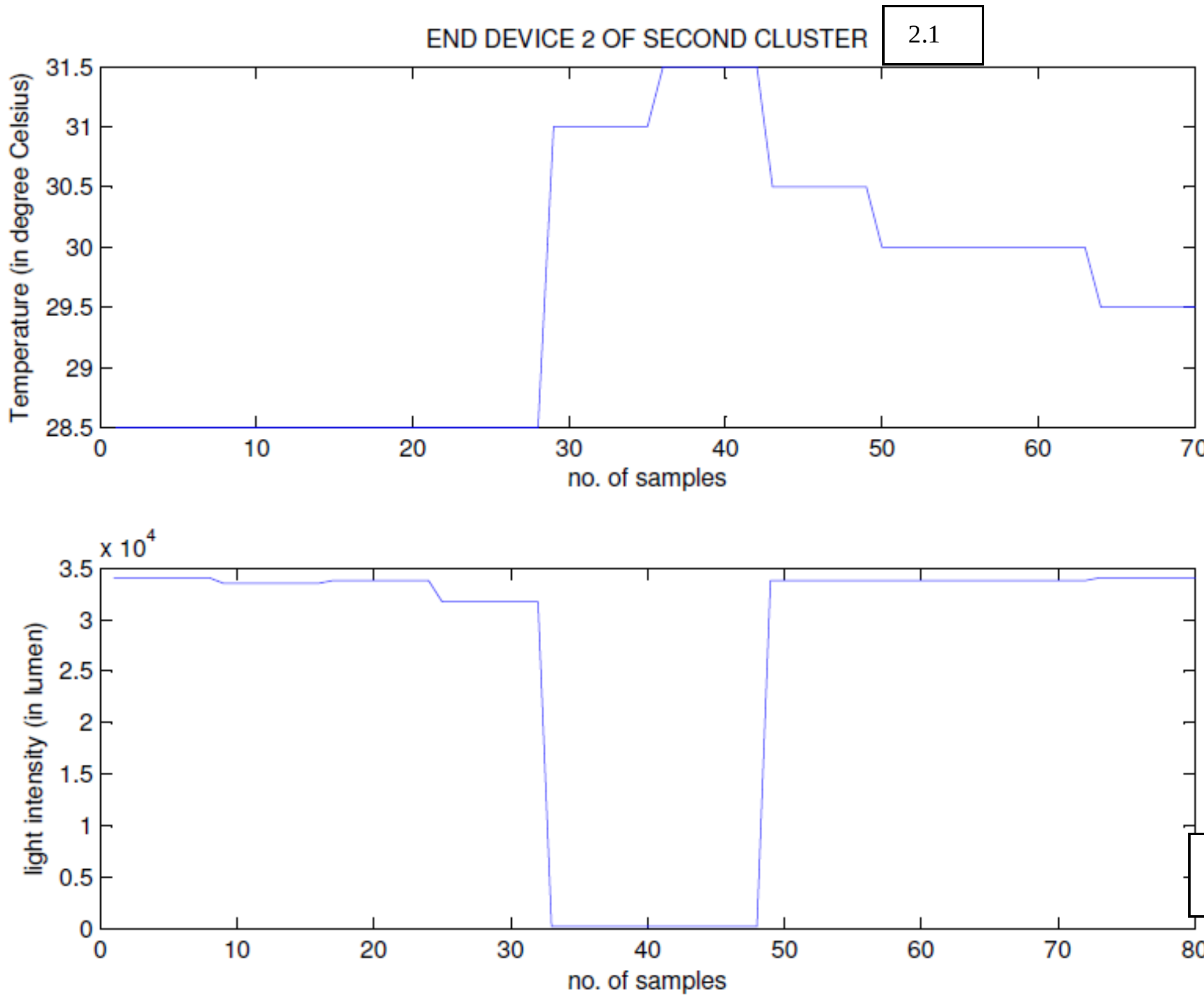

***Figure 4.6.6***

There is not much of variation in the readings because all nodes were placed in the same room. But important part is that the readings can be seen on the graphs.

# 5. Conclusions

I. We have successfully implemented a multi-hop network of sensor nodes which are capable of monitoring temperature & light intensity. The base station for this network is responsible for aggregation of data from all nodes and sending it to the main server via a gateway. The server then provides service to multiple clients.

II. Easily available components, proven network concepts, open source software and codes have been used to demonstrate our system can be made in low cost.

III. As most of the software made and used is open source further modifications can be made as per the requirements for the safety of mines.

IV. Though the work here has been mine centric the concept can open entirely new areas of business for telecom operators and web service providers, e.g. monitoring ones home while at work or on a long tour.

We believe that our project can render the workplace of miners safe from unpredictable situations and accidents which lead to loss of human lives and prosperity.

# 6. Proposal for Future work

We envision a fully fledged technical program leveraging state of art wireless sensor network technologies towards an intelligent and secure mining environment. It can render the workplace of miners safe from unpredictable situations and accidents which lead to loss of human lives and prosperity.

**We already have reached a level where we have shown the effectiveness of wireless sensor network with a small set of nodes. This work can be taken to a much bigger platform which can bring drastic changes in the present situation of mine safety.**

## 6.1 Extended features in our hardware

The hardware we have used presently gives the measure of temperature and light intensity which can be extended to measure several other parameters important from the safety point of view. Like concentration of oxygen, carbon monoxide nitrogen, and methane which can be done fabricating gas sensors like TGS-2611, TGS-2442, and TGS-2600; water level and moisture content can also be measured by using suitable sensors.

## 6.2 Improving the ruggedness and quality of the hardware

Although, the hardware which we are using can be further used to fabricate more sensors and other components but it may not work effectively in the actual atmosphere of mine. So, the quality of hardware may be improved to adapt in the harsh conditions keeping the architecture and the components same.

### 6.3 Mine Surveillance System

Extending its features, it can be taken to much higher level applications like automated alarm system so that people know in advance that conditions in the mines are dangerous and needs to be vacated.

### 6.4 Making the sensor nodes mobile

The nodes of the wireless sensor network can be made mobile i.e. they can be built on automated robots for easy access to various places.

### 6.5 Simulation of the Mines

Simulation of the mines can be done by software which will enable us to have virtual tour of the actual mine. So, that the workers are aware of the conditions in the mines before hand and hence can be well trained to negate the level of risk to their lives.

## Acknowledgements

1. Vividh Mishra, Senior Project Associate, *Department of Electrical Engineering, Indian Institute of Technology, Kanpur – 208016, India*. *E-mail:* mishravividh@gmail.com

2. Pankaj Prashant, B.Tech. III Yr, Department of Mining Engineering, Indian School of Mines University, Dhanbad – 826004, India. E-mail: pankaj.prashant1@gmail.com

3. Nalin Gupta, B. Tech. III Yr , Department of Electronics & Instrumentation, Indian School of Mines University, Dhanbad – 826004,India.
E-mail: nalingupta@ismu.ac.in

# Annexure 1

## ➢ *Previous Statistics of Coal mine disasters*

| ***Number of people*** | ***Place &situation*** |
|---|---|
| 1,549 | Benxihu Colliery explosion, (China, 1942) |
| 1,099 | Courrières mine disaster (Courrières, France, 1903) |
| 687 | coal mine (Mitsubishi Hojo, Kyūshū, Japan, 15 December 1914) |
| 682 | coal mine Laobaidong colliery coal dust explosion, (Datong China, 9 May 1960) |
| 472 | coal mine (Wankie, Rhodesia, 1972) |
| 458 | coal mine (Mitsui Miike, Omuta, Kyūshū, Japan, 9 November 1963) |
| 439 | Senghenydd Colliery Disaster (Senghenydd, Wales, 1913) |
| 437 | coal mine (Coalbrook, South Africa, 1960) |
| 422 | coal mine (New Yubari, Yubari, Hokkaidō, Japan, 28 November 1914) |
| 405 | coal mine (Bergkamen, West Germany, 1946) |
| 388 | Oaks Colliery, (Barnsley, England, 1866) |
| 376 | coal mine (Onoura, Kirino, Kyūshū, Japan, 21 December 1917) |
| 375 | coal mine (Bihar, India, 1965) |
| 372 | Chasnala mining disaster,Sudamdih Colliery (Dhanbad, India, 1975) |
| 365 | coal mine (Hokoku, Itoda, Kyūshū, Japan, 20 July 1907) |
| 362 | coal mine (Monongah, West Virginia, 1907) |
| 344 | Pretoria Pit Disaster, (Westhoughton, England, 1910) |
| 298 | coal mine (Saarland, West Germany, 1962) |
| 266 | Gresford Disaster (Gresford, Wales, 1934) |
| 263 | coal mine (Dawson, New Mexico, 1913) |
| 263 | coal mine (Incirharmani, Kozlu[disambiguation needed], Zonguldak, Turkey, March 3, 1992) |
| 262 | coal mine (Marcinelle, Belgium, 1956) |
| 259 | coal mine (Cherry, Illinois, 1909) |
| 257 | coal mine (Grundy, Virginia, 1937) |
| 243 | coal mine (Onoura, Kirino, Kyūshū, Japan, 5 August 1909) |
| 239 | coal mine (Jacobs Creek, Pennsylvania, 1907) |
| 236 | coal mine (Chikuho Yamano[disambiguation needed], Kyūshū, Japan, 1 June 1965) |
| 235 | coal mine (Larisch, Karviná, Czech Republic, June 14, 1894) |
| 214 | coal mine (Sunjiawan, Fuxin, Liaoning, China, 15 February 2005) |
| 210 | coal mine (Hokoku, Itoda, Kyūshū, Japan, 15 June 1899) |
| 200 | coal mine (Scofield, Utah, 1900) |
| 200 | coal mine (Mina Rosita Vieja disaster, San Juan de Sabinas, Coahuila, Mexico, 27 February 1908) |
| 189 | coal mine (Hillcrest mine, Canada, 1914) |
| 183 | coal mine (Chosei, Ube, Japan, 3 February 1942) |
| 181 | coal mine with flooding (Huayuan[disambiguation needed], Xintai, Shandong, China, August 17, 2007) |
| 180 | coal mine (Tuzla, Bosnia and Herzegovina, 1990) |
| 177 | coal mine (Mitsubishi Bibai, Bibai, Hokkaidō, Japan,1941) |
| 166 | coal mine (Chenjiashan, Tongchuan, Shaanxi, China, 28 November 2004) |
| 159 | coal mine (Muchonggou, Shuicheng, Guizhou, China, 26 September 2000) |
| 155 | coal mine (Minnie Pit, Podmore Hall, Halmer End, Staffordshire, UK, 12 January 1918) |
| 153 | coal mine (Mina de Barroteran, Coahuila, Mexico, March 31, 1969) |
| 151 | coal mine (Hausdorf, Germany currently Jugów, Poland, July 9, 1930) |
| 150 | Nanaimo mine explosion (Canada, 1887) |
| 148 | coal mine (Daping, Tongchuan[disambiguation needed], Henan, China, 20 October 2004) |
| 147 | coal mine (Sanjiaohe, Hongdong, Shanxi, China, April 21, 1991) |
| 134 | coal mine (Uchigo, Iwaki[disambiguation needed], Japan, March 27, 1927) |
| 134 | coal mine (Dongfeng, Qitaihe, Heilongjiang, China, 27 November 2005) |
| 125 | coal mine (Springhill, Nova Scotia, Canada, 21 February 1891) |
| 125 | coal mine (Amaga, Angelopolis, Antioquia, Colombia, July 15, 1977) |
| 124 | coal mine (Chengzihe, Jixi, Heilongjiang, China, 20 June 2002) |
| 123 | coal mine (Daxing, Xingning, Guangdong, China, 6 August 2005) |
| 119 | coal mine (Orient #2 Mine, West Frankfort, IL, December 25, 1951) |
| 111 | coal mine (Centralia Mine Disaster, Centralia, IL, March 25, 1947) |
| 109 | coal mine (Everettville, WV, April 30, 1927) |
| 108 | coal mine (Ulyanovskaya, Novokuznetsk, Kuzbass Siberia, Russia, March 19, 2007) |
| 108 | coal mine (Dukla, Ostrava, Czech Republic, July 14, 1961) |
| 108 | coal mine (Larisch, Karviná, Czech Republic, March 5, 1885) |
| 105 | coal mine (Ruizhiyuan, Linfen, Shanxi, China, 5 December 2007) |
| 101 | coal mine (methane explosion) (Zasyadko, Donetsk, Ukraine, 18 |
| | November 2007) |
| 100 | coal mine (Armutçuk, Ereğli, Zonguldak, Turkey, March 8, 1983) |
| 94 | coal mine (Haishan, Taipei, Taiwan, December 5, 1984) |
| 93 | coal mine (Hokutan Yubari, Yubari, Hokkaidō, Japan, 16 October 1981) |
| 92 | coal mine (Rudnici, Aleksinac, Nisava District, Serbia, November 18, 1989) |
| 92 | coal mine (Gangzi, Xuzhou, Jiangsu, China, July 22, 2001) |
| 91 | coal mine (Dongcun, Datong, Shanxi, China, 27 November 1996) |
| 89 | coal mine (Xingsheng, Pingdingshan, Henan, China, March 11, 1997) |
| 86 | coal mine (Luling coal mine, Hefei, Anhui, China, 13 May 2003) |
| 83 | coal mine (Mitsui Miike, Omuta, Kyūshū, Japan, 18 January 1984) |
| 83 | coal mine (Shenlong, Fukang, Xinjiang Uygur, China, 13 July 2005) |
| 82 | coal mine (Barakova, Krasnodon, Ukraine, 11 March 2000) |
| 74 | coal mine (Springhill mining disaster, Springhill, Nova Scotia, Canada, 23 October 1958) |
| 71 | coal mine (Ueda Kamikiyo, Kawara, Kyūshū, Japan, 9 March 1961) |
| 71 | coal mine (Trimdon Grange], Trimdon Grange, County Durham, England 16 February 1882} |
| 68 | coal mine (Zyryanouskaya, Novokuznetsk, Kuzbass Siberia, Russia, December 2, 1997) |
| 65 | coal mine (Pluto, Záluží, Czech Republic, September 4, 1981) |
| 64 | coal mine (Hlobane, Vryheid, Kwa-Zulu Natal, South Africa, September 3, 1983) |
| 63 | coal mine (Skochynsky, Donetsk, Ukraine, 4 April 1998) |
| 60 | coal mine (Beilongfeng, Fushun, Liaoning, China, May 28, 1997) |
| 59 | coal mine (Xishui mine, Shanxi province, China, 20 March 2005) |
| 53 | coal mine (Secunda, Mpumalanga, East Transvaal, South Africa, May 13, 1993) |
| 52 | coal mine (Zasyadka, Donetsk, Ukraine, 19 August 2001) |
| 45 | coal mine (Abertillery, Blaenau Gwent, Wales, 28 June 1960) |
| 43 | coal mine (Muchonggou, Shuicheng, Guizhou, China, 24 February 2003) |
| 39 | coal mine (Springhill, Nova Scotia, Canada, 1 November 1956) |
| 36 | coal mine (Donbass, Donetsk, Ukraine, 19 July 2004) |
| 30 | coal mine (Abaiskaya, Abai[*disambiguation needed*], Karaganda Oblast, Kazakhstan, January 11, 2008) |
| 29 | Upper Big Branch mine explosion (Montcoal, West Virginia, 5 April 2010) |
| 29 | coal mine (Nordegg, Alberta, 31 October 1941) |
| 26 | coal mine (Westray Mine, Nova Scotia, Canada, May 9, 1992) |
| 24 | Nanshan Colliery disaster (Shanxi Province, China, 13 November 2006) |
| 23 | methane explosion in Halemba coal mine (Ruda Śląska Poland 21 November 2006) |
| 21 | coal mine (Kings Bay, Svalbard, Norway, November 5, 1962)no:Kings Bay-saken |
| 20 | 2009 Handlová mine blast (Handlová, Slovakia, August 10, 2009) |
| 17 | coal mine (Mine No. 9, Sturgis, Union County, Kentucky), 8, June 1925 |
| 16 | coal mine (Ningxia Hui, China, 16 October 2008) |
| 12 | Sago Mine Disaster (Sago, West Virginia, 2 January 2006 |

In reality, there have been many more such accidents especially in developing nations like India but were not being documented. So, the death toll is much more then we see here.